# OSMnx: New Methods for Acquiring, Constructing, Analyzing, and Visualizing Complex Street Networks

Geoff Boeing
Email: gboeing@berkeley.edu
Department of City and Regional Planning
University of California, Berkeley
May 2017

Abstract

Urban scholars have studied street networks in various ways, but there are data availability and consistency limitations to the current urban planning/street network analysis literature. To address these challenges, this article presents OSMnx, a new tool to make the collection of data and creation and analysis of street networks simple, consistent, automatable and sound from the perspectives of graph theory, transportation, and urban design. OSMnx contributes five significant capabilities for researchers and practitioners: first, the automated downloading of political boundaries and building footprints; second, the tailored and automated downloading and constructing of street network data from OpenStreetMap; third, the algorithmic correction of network topology; fourth, the ability to save street networks to disk as shapefiles, GraphML, or SVG files; and fifth, the ability to analyze street networks, including calculating routes, projecting and visualizing networks, and calculating metric and topological measures. These measures include those common in urban design and transportation studies, as well as advanced measures of the structure and topology of the network. Finally, this article presents a simple case study using OSMnx to construct and analyze street networks in Portland, Oregon.

# 1. Introduction

Urban scholars and planners have studied street networks in numerous ways. Some studies focus on the urban form (e.g., Southworth and Ben-Joseph 1997; Strano et al. 2013), others on transportation (e.g., Marshall and Garrick 2010; Parthasarathi et al. 2013), and others on the topology, complexity, and resilience of street networks (e.g., Jiang and Claramunt 2004; Porta et al. 2006). This article argues that current limitations of data availability, consistency, and technology have made researchers' work gratuitously difficult. In turn, this empirical literature often suffers from four shortcomings which this article examines: small sample sizes, excessive network simplification, difficult reproducibility, and the lack of consistent, easy-to-use research tools. These shortcomings are by no means fatal, but their presence limits the scalability, generalizability, and interpretability of empirical street network research.

To address these challenges, this article presents OSMnx, a new tool that easily downloads and analyzes street networks for anywhere in the world. OSMnx contributes five primary capabilities for researchers and practitioners. First, it enables automated and on-demand downloading of political boundary geometries, building footprints, and elevations. Second, it can automate and customize the downloading of street networks from OpenStreetMap and construct them into multidigraphs. Third, it can correct and simplify network topology. Fourth, it can save/load street networks to/from disk in various file formats. Fifth and finally, OSMnx has built-in functions to analyze street networks, calculate routes, project and visualize networks, and quickly and consistently calculate various metric and topological measures. These measures include those common in urban design and transportation studies, as well as advanced measures of the structure and topology of the network.

This article is organized as follows. First, it introduces the background of networks, street network analysis and representation, and the current landscape of tools for this type of research. Then it discusses shortcomings and current challenges, situated in the empirical literature. Next, it introduces OSMnx and its methodological contributions. Finally, it presents a simple illustrative case study using OSMnx to construct and analyze street networks in Portland, Oregon, before concluding with a discussion.

# 2. Background

Street network analysis has been central to network science since its nascence: its mathematical foundation, graph theory, was born in the 18th century when Leonhard Euler presented his famous Seven Bridges of Königsberg problem. Here we briefly trace the fundamentals of modern street network research from graphs to networks to the present landscape of research toolkits, in order to identify current limitations.

## 2.1. Graphs and networks

Network science is built upon the foundation of graph theory, a branch of discrete mathematics. A *graph* is an abstract representation of a set of elements and the connections between them (Trudeau 1994). The elements are interchangeably called vertices or nodes, and the connections between them are called links or edges. For consistency, this article uses the terms *nodes* and *edges*. The number of nodes in the graph (called the *degree* of the graph) is commonly represented as $n$ and the number of edges as $m$. Two nodes are *adjacent* if an edge connects them, two edges are adjacent if they share the same node, and a node and an edge are *incident* if the edge connects the node to another node. A node's *degree* is the number of edges incident to the node, and its *neighbors* are all those nodes to which the node is connected by edges.

An *undirected* graph's edges point mutually in both directions, but a *directed* graph, or digraph, has directed edges (i.e., edge *uv* points from node *u* to node *v*, but there is not necessarily a reciprocal edge *vu*). A *self-loop* is an edge that connects a single node to itself. Graphs can also have parallel (i.e., multiple) edges between the same two nodes. Such graphs are called *multigraphs*, or *multidigraphs* if they are directed.

An undirected graph is *connected* if each of its nodes can be reached from any other node. A digraph is *weakly connected* if the undirected representation of the graph is connected, and *strongly connected* if each of its nodes can be reached from any other node. A *path* is an ordered sequence of edges that connects some ordered sequence of nodes. Two paths are *internally node-disjoint* if they have no nodes in common, besides end points. A *weighted* graph's edges have a weight attribute to quantify some value, such as importance or impedance, between connected nodes. The *distance* between two nodes is the number of edges in the path between them, while the *weighted distance* is the sum of the weight attributes of the edges in the path.

While a graph is an abstract mathematical representation of elements and their connections, a *network* may be thought of as a real-world graph. Networks inherit the terminology of graph theory. Familiar examples include social networks (where the nodes are humans and the edges are their interpersonal relationships) and the World Wide Web (where the nodes are web pages and the edges are hyperlinks that point from one to another). A *complex* network is one with a nontrivial *topology* (the configuration and structure of its nodes and edges) – that is, the topology is neither fully regular nor fully random. Most large real-world networks are complex (Newman 2010). Of particular interest to this study are *complex spatial networks* – that is, complex networks with nodes and/or edges embedded in space (O'Sullivan 2014). A street network is an example of a complex spatial network with both nodes and edges embedded in space, as are railways, power grids, and water and sewage networks (Barthélemy 2011).

## 2.2. Representation of street networks

A spatial network is *planar* if it can be represented in two dimensions with its edges intersecting only at nodes. A street network, for instance, *may* be planar (particularly at certain small scales), but most street networks are non-planar due to grade-separated expressways, overpasses, bridges, and tunnels. Despite this, most quantitative studies of urban street networks represent them as planar (e.g., Buhl et al. 2006; Cardillo et al. 2006; Barthélemy and Flammini 2008; Masucci et al. 2009; Strano et al. 2013) for tractability because bridges and tunnels are reasonably uncommon (in certain places) – thus the networks are *approximately* planar. However, this over-simplification to planarity for tractability may be unnecessary and can cause analytical problems, as we discuss shortly.

The street networks discussed so far are *primal*: the graphs represent intersections as nodes and street segments as edges. In contrast, a *dual graph* (namely, the edge-to-node dual graph, also called the *line graph*) inverts this topology: it represents a city's streets as nodes and intersections as edges (Porta et al. 2006). Such a representation may seem a bit odd but provides certain advantages in analyzing the network topology based on named streets (Crucitti et al. 2006). Dual graphs form the foundation of space syntax, a method of analyzing urban networks and configuration via axial street lines and the depth from one edge to others (Hillier et al. 1976; cf. Ratti 2004). Jiang and Claramunt (2002) integrate an adapted space syntax – compensating for difficulties with axial lines – into computational GIS. Space syntax has formed the basis of various other adapted approaches to analytical urban design (e.g., Karimi 2012).

This present article, however, focuses on primal graphs because they retain all the geographic, spatial, metric information essential to urban form and design that dual representations discard: all the geographic, experiential traits of the street (such as its length, shape, circuity, width, etc.) are lost in a dual graph. A primal graph, by contrast, can faithfully represent all the spatial characteristics of a street. Primal may be a better approach for analyzing spatial networks when geography matters, because the physical space underlying the network contains relevant information that cannot exist in the network's topology alone (Ratti 2004).

### 2.3. Street network analysis

Street networks – considered here as primal, non-planar, weighted multidigraphs with self-loops – can be characterized and described by metric and topological measures. Extended definitions and algorithms can be found in, e.g., Newman (2010) and Barthélemy (2011).

*Metric structure* can be measured in terms of length and area and represents common transportation/design variables (e.g., Cervero and Kockelman 1997; Ewing and Cervero 2010). *Average street length*, the mean edge length (in spatial units such as meters) in the undirected representation of the graph, serves as a linear proxy for block size and indicates how fine-grained or coarse-grained the network is. *Node density* is the number of nodes divided by the area covered by the network. *Intersection density* is the node density of the set of nodes with more than one street emanating from them (thus excluding dead-ends). The *edge density* is the sum of all edge lengths divided by the area, and the physical *street density* is the sum of all edges in the undirected representation of the graph divided by the area. These density measures all provide further indication of how fine-grained the network is. Finally, the *average circuity* divides the sum of all edge lengths by the sum of the great-circle distances between the nodes incident to each edge (cf. Giacomin and Levinson 2015). This is the average ratio between an edge length and the straight-line distance between the two nodes it links.

The *eccentricity* of a node is the maximum of the shortest-path weighted distances between it and each other node in the network. This represents how far the node is from the node that is furthest from it. The *diameter* of a network is the maximum eccentricity of any node in the network and the *radius* of a network is the minimum eccentricity of any node in the network. The *center* of a network is the node or set of nodes whose eccentricity equals the radius and the *periphery* of a network is the node or set of nodes whose eccentricity equals the diameter. When weighted by length, these distances indicate network size and shape in units such as meters.

*Topological measures* of street network structure indicate the configuration, connectedness, and robustness of the network – and how these characteristics are distributed. The *average node degree*, or mean number of edges incident to each node, quantifies how well the nodes are connected on average. Similarly, but more concretely, the *average streets per node* measures the mean number of physical streets (i.e., edges in the undirected representation of the graph) that emanate from each intersection and dead-end. This adapts the average node degree for physical form rather than directed circulation. The statistical and spatial distributions of number of streets per intersection characterize the type, prevalence, and dispersion of intersection connectedness in the network.

*Connectivity* measures the minimum number of nodes or edges that must be removed from a connected graph to disconnect it. This is a measure of resilience as complex networks with high connectivity provide more routing choices to agents and are more robust against failure. However, node and edge connectivity is less useful for approximately planar networks like street networks: *most* street networks will have connectivity of 1, because the presence of a single dead-end means the removal of just one node or edge can disconnect the network. More usefully, the *average node connectivity* of a network – the mean number of internally node-disjoint paths between each pair of nodes in the graph – represents the expected number of nodes that must be removed to disconnect a randomly selected pair of non-adjacent nodes (Beineke et al. 2002).

As O'Sullivan (2014) discusses, network distances, degrees, and connectivity are significantly constrained by spatial embeddedness and approximate planarity. Other measures of connectedness – such as intersection density, node degree distribution, and centrality/clustering (discussed below) – may better capture the nature of a street network's connectedness than node or edge connectivity can. Networks with low connectedness may have multiple single points of failure, leaving the system particularly vulnerable to perturbation. This can be seen in urban design through permeability and choke points: if circulation is forced through single points of failure, traffic jams can ensue and circulation networks can fail. Connectedness has also been linked to street network pedestrian volume (Hajrashouliha and Yin 2015).

Clustering measures also reveal topological structure and its distribution. The *clustering coefficient* of a node is the ratio of the number of edges between its neighbors to the maximum possible number of edges that could exist between these neighbors. The *weighted clustering coefficient* weights this ratio by edge length and the *average clustering coefficient* is the mean of the clustering coefficients of all the nodes in the network. These measure connectedness and complexity by how thoroughly the neighborhood of some node is linked together. Jiang and Claramunt (2004) extend this coefficient to neighborhoods within an arbitrary distance to make it more applicable to urban street networks.

Centrality indicates the importance of nodes in a network. *Betweenness centrality* evaluates the number of shortest paths that pass through each node or edge (Barthélemy 2004). The maximum betweenness centrality in a network specifies the proportion of shortest paths that pass through the most important node/edge. This is an indicator of resilience: networks with a high maximum betweenness centrality are more prone to failure or inefficiency should this single choke point fail. The average betweenness centrality is the mean of all the betweenness centralities in the network (Barthélemy 2011). Barthélemy et al. (2013) use betweenness centrality to identify top-down interventions versus bottom-up self-organization and evolution of Paris's urban fabric.

*Closeness centrality* represents, for each node, the reciprocal of the sum of the distance from this node to all others in the network: that is, nodes rank as more central if they are on average closer to all other nodes. Finally, *PageRank* – the algorithm Google uses to rank web pages – is a variant of network centrality, namely eigenvector centrality (Brin and Page 1998). PageRank ranks nodes based on the structure of incoming links and the rank of the source node, and may also be applied to street networks (Agryzkov et al. 2012; Chin and Wen 2015).

## 2.4. Current tool landscape

Several tools exist to study street networks. ESRI provides an ArcGIS Network Analyst extension, for which Sevtsuk and Mekonnen (2012) developed the Urban Network Analysis Toolkit plug-in. QGIS, an open-source alternative, also provides limited capabilities through built-in plug-ins. GIS tools generally provide very few network analysis capabilities, such as shortest path calculations. In contrast, network analysis software – such as Gephi, igraph, and graph-tool – provides minimal GIS functionality to study spatial networks. Pandana is a Python package that performs accessibility queries over a spatial network, but does not support other graph-theoretic network analyses (Foti 2014). NetworkX is a Python package for general network analysis, developed by researchers at Los Alamos National Laboratory. It is free, open-source, and able to analyze networks with millions of nodes and edges (Hagberg and Conway 2010).

Street network data comes from many sources – including city, state, and national data repositories – and typically in shapefile format. Expensive proprietary data sources such as HERE NAVSTREETS and TomTom MultiNet also exist. In the US, the census bureau provides free TIGER/Line (Topologically Integrated Geographic Encoding and Referencing) shapefiles of geographic data such as cities, census tracts, roads, buildings, and certain natural features. However, TIGER/Line roads shapefiles suffer from inaccuracies (Frizelle et al. 2009), contain quite coarse-grained classifiers (e.g., classifying parking lots as alleys), and topologically depict bollarded intersections as through-streets, which problematizes routing. Furthermore, there is no central repository of worldwide street network data, which can be inconsistent, difficult, or impossible for researchers to obtain in many countries.

OpenStreetMap – a collaborative mapping project that provides a free and publicly editable map of the world – has emerged in recent years as a major player both for mapping and for acquiring spatial data (Corcoran et al. 2013; Jokar Arsanjani 2015). Inspired by Wikipedia's mass-collaboration model, the project started in 2004 and has grown to over two million users today. Its data quality is generally quite high (Girres and Touya 2010; Haklay 2010; Barron et al. 2014) and although data coverage varies worldwide, it is generally good when compared to corresponding estimates from the CIA World Factbook (Maron 2015). In the US, OpenStreetMap imported the 2005 TIGER/Line roads in 2007 as a foundational data source (Zielstra et al. 2013). Since then, numerous corrections and improvements have been made. But more importantly, many additions have been made beyond what TIGER/Line captures, including pedestrian paths through parks, passageways between buildings, bike lanes and routes, and richer attribute data describing the characteristics of features, such as finer-grained codes for classifying arterial roads, collector streets, residential streets, alleys, parking lots, etc.

There are several methods of acquiring street network data from OpenStreetMap. First, OpenStreetMap provides an API, called Overpass, which can be queried programmatically to retrieve any data in the database: streets and otherwise. However, its usage and syntax are notoriously challenging and several services have sprung up to simplify the process. Mapzen extracts chunks of OpenStreetMap data constrained to bounding boxes around 200 metropolitan areas worldwide. They also provide custom extracts, which can take up to an hour to run. Mapzen works well for simple bounding boxes around popular cities, but otherwise does not provide an easily scalable or customizable solution. Geofabrik similarly provides data extracts, generally at the national or sub-national scale, but provides shapefiles as a paid service.

Finally, GISF2E is a tool (compatible with ArcGIS and an outdated version of Python) that can convert shapefiles (such as Mapzen or Geofabrik extracts) into graph-theoretic network data sets (Karduni et al. 2016). Its creators provide processed shapefiles for several cities online but with some limitations. While GISF2E shapefiles' roads have a flag denoting one-way streets, it discards *to* and *from* nodes, thus making it unclear in which direction the one-way goes. It also can treat nodes inconsistently due to arbitrary break points between OpenStreetMap IDs or line digitization. OpenStreetMap IDs *sometimes* map 1-to-1 with a named street, but other times a named street might comprise multiple OpenStreetMap IDs. Further, some streets have arbitrary "nodes" in the middle of a segment because the OpenStreetMap ID is different on either side.

## 2.5. Research problem

Due to the preceding limitations of street network data availability, consistency, and technology, the empirical literature often suffers from four shortcomings. First, the sample sizes in cross-sectional studies tend to be quite small due to clear challenges in acquiring large data sets. Most cross-sectional studies tend to analyze somewhere between 10 and 50 or so networks for tractability at the city or neighborhood scale (e.g., Buhl et al. 2006; Cardillo et al. 2006; Jiang 2007; Marshall and Garrick 2010; Strano et al. 2013; Giacomin and Levinson 2015). Acquiring and assembling large numbers of street networks consistently from data sources spread across various governmental entities can be extremely difficult and time-consuming. However, small sample sizes can limit the representativeness and reliability of findings.

Second, studies usually simplify the representation of the street network to a planar or undirected graph for tractability (e.g., Buhl et al. 2006; Cardillo et al. 2006; Barthélemy and Flammini 2008; Masucci et al. 2009). Typically, researchers assemble street networks into some sort of graph-theoretic object from GIS data, for instance by splitting the centerlines of all the streets in a study area wherever they cross in two dimensions. These split lines become edges and the splitting points become nodes. However, this method presumes a planar graph: bridges and tunnels become splitting points (and thus nodes) even if the streets do not actually intersect in three dimensions. Unless the street network is truly planar, a planar simplification is thus a less-than-ideal representation that potentially yields inaccurate metrics, underestimates the lengths of edges, and overestimates the number of nodes. This may reasonably represent a street network in a European medieval city center, but poorly represents the street network in a city like Los Angeles with numerous grade-separated expressways, bridges, and tunnels in a truly non-planar network. Karduni et al. (2016) suggest the importance of using GIS attribute data to identify such non-planar features to create a correct topology.

The third problem is replicability. The dozens of decisions that go into analysis – such as spatial extents, topological simplification and correction, definitions of nodes and edges, etc. – are often ad hoc or only partially reported, making reproducibility challenging. Some studies gloss over the precise details of how their street networks were constructed (frequently due to some combination of methodological complexity and journal word limits), yet numerous unreported decisions had to be made in the process. For example, various studies examine cities out to the urban periphery, but do not explain precisely how and where this periphery is defined (e.g., Strano et al. 2013). Some studies do not report if their networks are directed or undirected (e.g., Porta et al. 2006; Strano et al. 2013). Directedness may not matter for pedestrian studies but substantially impacts the interpretation of various network statistics when directedness does matter (e.g., for drivable networks). Further, what are edges in the street network? Drivable streets? Pedestrian paths? What is a node in the street network? Is it where at least two different named streets come together? Does it denote any junction of routes? What about dead-ends? Different studies make different but perfectly valid decisions with these various questions (e.g., Frizelle et al. 2009; Marshall and Garrick 2010; Sevtsuk and Mekonnen 2012). Their definitions impact how we interpret various calculated features like degrees or intersection densities, and any research design decisions that go unreported can problematize replicability, interpretation, and generalizability.

Fourth, as discussed, the current landscape of tools and methods offers no ideal technique that balances usability, customizability, reproducibility, and scalability in acquiring, constructing, and analyzing network data. Taken together, these limitations make street network researchers' work difficult and can circumscribe the conclusions that may be drawn from the effort.

# 3. OSMnx: Functionality and comparison to existing tools

To address these challenges, this article presents a new tool to make the collection of data and creation and analysis of street networks simple, consistent, automatable, and sound from the perspectives of graph theory, transportation, and urban design. OSMnx is a free, open-source Python package that downloads political/administrative boundary geometries, building footprints, and street networks from OpenStreetMap. It enables researchers to easily construct, project, visualize, and analyze non-planar complex street networks consistently by constructing a city's or neighborhood's walking, driving, or biking network with a single line of Python code – including node elevations and street grades. Python was chosen because the language is popular, easy for beginners, powerful, fast, free, and open-source. OSMnx is built on top of Python's NetworkX, matplotlib, and geopandas libraries for rich network analytic capabilities, beautiful and simple visualizations, and fast spatial queries with R-tree indexing. This section discusses OSMnx's primary features in order.

## 3.1. Acquire political boundaries and building footprints

To acquire place boundary GIS data, researchers typically must locate and download shapefiles online. However, bulk or automated acquisition (such as that required to analyze hundreds or thousands of separate geographies) requires clicking through numerous web pages to download shapefiles one at a time. With OSMnx, one can download place geometries from OpenStreetMap for anywhere in the world with a single line of Python code, and project and visualize in one more line of code (Figure 1). To project, OSMnx calculates UTM zones algorithmically based on the centroid of the geometry. One can easily acquire geometries for multiple place types, such as neighborhoods, boroughs, counties, states, or nations – any place geometry available in OpenStreetMap. Or, one can pass multiple places in a single query to construct multiple features of cities, states, counties, nations, or any other geographic entities, and the results can be saved as shapefiles to a hard drive. Similarly, building footprints can be retrieved for anywhere that OpenStreetMap has such data.

## 3.2. Download and construct street networks

The primary use of OSMnx is the easy downloading and construction of street networks. To acquire street network GIS data, one must typically track down TIGER/Line roads from the US census bureau – or individual data sets from other countries or cities, as TIGER/Line provides no street network data for geographies outside the US – then convert the data to graph-theoretic objects. However, this becomes preventively burdensome for large numbers of separate street networks as it does not lend itself to bulk, automated analysis. Further, it ignores informal paths and pedestrian circulation that TIGER/Line lacks. In contrast, OSMnx handles all of these use cases.

OSMnx downloads street network data and builds topologically-corrected street networks, projects and plots the networks, and saves them as SVGs, GraphML files, or shapefiles for later use. The street networks are primal, non-planar, weighted multidigraphs with self-loops and they preserve one-way directionality. One can download a street network by providing OSMnx any of the following queries (Figure 2):

- bounding box
- latitude-longitude point plus distance in meters (either a distance along the network or a distance in each cardinal direction from the point)
- address plus distance in meters (either a distance along the network or a distance in each cardinal direction from the point)
- polygon of the desired street network's boundaries
- place name or list of place names

One can also specify different network types to clarify what is an edge in the network:

- *drive*: get drivable public streets (but not service roads)
- *drive_service*: get drivable public streets including service roads
- *walk*: get all streets and paths that pedestrians can use (this network type ignores one-way directionality by always connecting adjacent nodes with reciprocal directed edges)
- *bike*: get all streets and paths that cyclists can use
- *all*: download all (non-private) OpenStreetMap streets and paths
- *all_private*: download all OpenStreetMap streets and paths, including private-access

When passed a place name (as in Figure 3), OSMnx geocodes it using OpenStreetMap's Nominatim API and constructs a polygon from its boundary geometry. It then buffers this polygon by 500 meters and downloads the street network data within this buffer from OpenStreetMap's Overpass API. Next it constructs the street network from this data. For one-way streets, directed edges are added from the origin node to the destination node. For bidirectional streets, reciprocal directed edges are added in both directions between nodes. Then OSMnx corrects the topology (discussed below), calculates accurate degrees and node types, then truncates the network to the original, desired polygon. This ensures that intersections are not considered dead-ends simply because an incident edge connects to a node outside the desired polygon. Finally, node elevation data and street grades may be added to the network from the Google Maps Elevation API in one more line of code.

Researchers can request a street network within a borough, county, state, or other geographic entity – or pass a list of places, such as several neighboring cities, to create a unified street network within the union of their geometries. In general, US roads GIS data is fairly easy to acquire, thanks to TIGER/Line shapefiles. OSMnx makes it *easier* by downloading it and turning it into a graph with a single line of code, and *better* by supplementing it with all the additional data (both attributes and non-road routes) from OpenStreetMap. With OSMnx, researchers can just as easily acquire street networks from anywhere in the world – including places in the Global South where such data might otherwise be inconsistent or difficult to acquire (Figure 4).

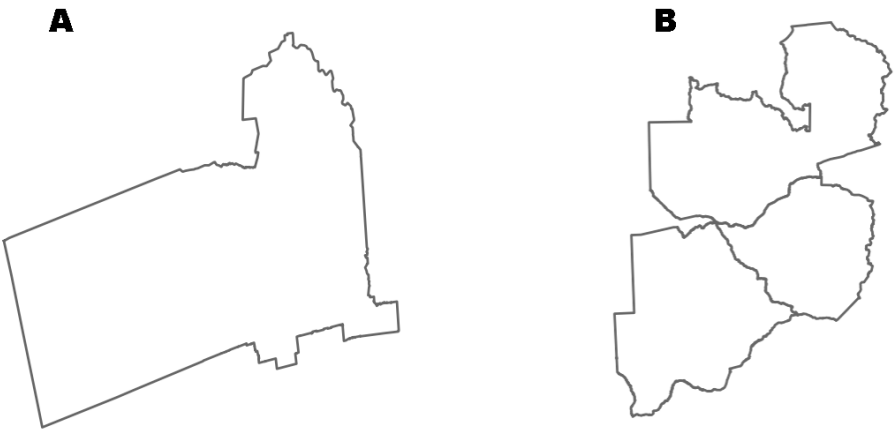

Figure 1. Administrative boundary vector geometries retrieved for A) the city of Berkeley, California and B) the nations of Zambia, Zimbabwe, and Botswana.

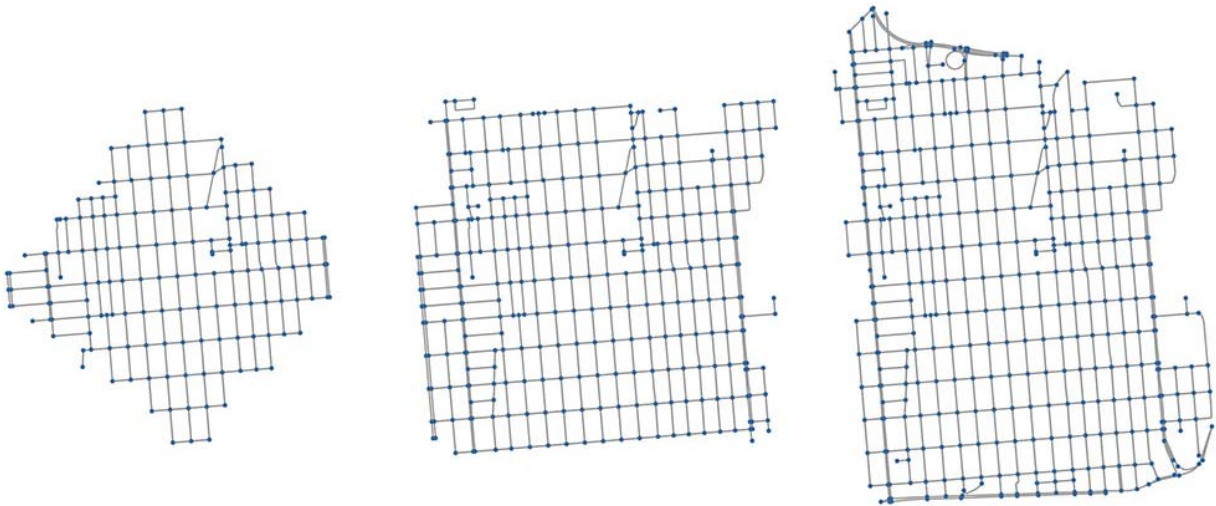

Figure 2. Three graphs from the same street network at the same scale, created by address and network distance (left), bounding box (center), and neighborhood polygon (right).

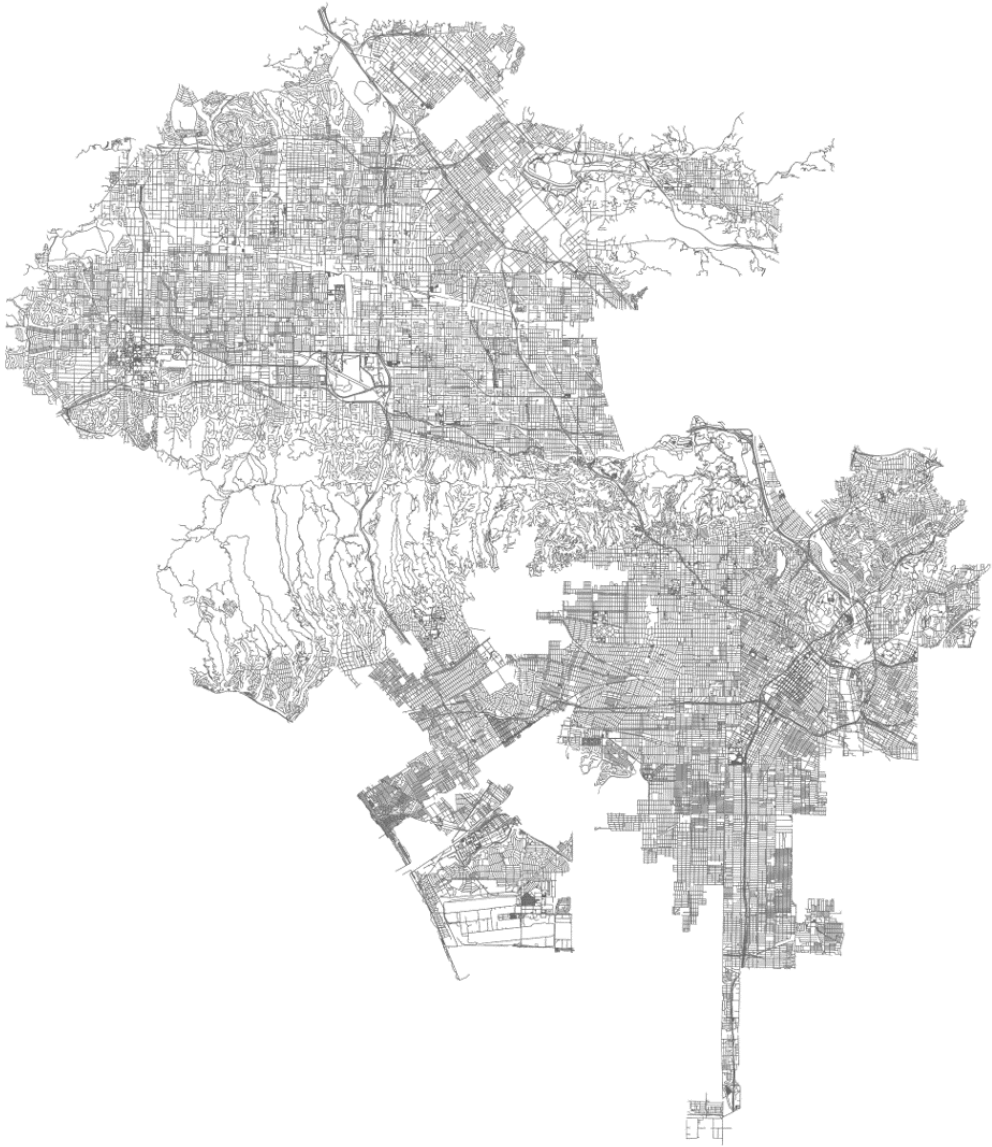

Figure 3. The drivable street network for municipal Los Angeles, created by simply passing the query phrase "Los Angeles, CA, USA" into OSMnx.

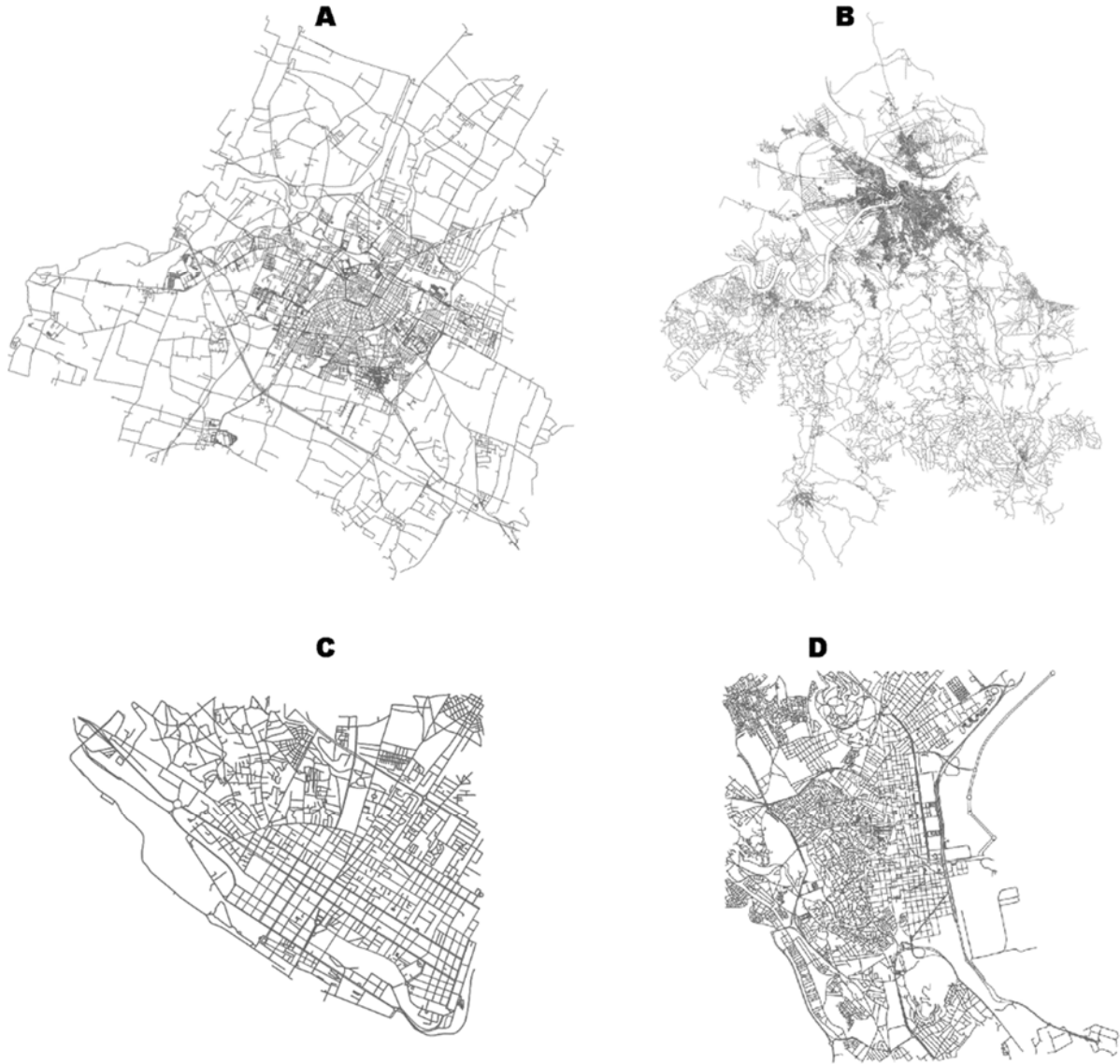

Figure 4. OSMnx street networks automatically downloaded and visualized for A) Modena, Italy, B) Belgrade, Serbia, C) central Maputo, Mozambique, and D) central Tunis, Tunisia.

### 3.3. Correct and simplify network topology

OSMnx performs topological correction and simplification automatically under the hood, but it is useful to inspect how it works. Simplification is essential for a correct topology because OpenStreetMap nodes are inconsistent: they comprise intersections, dead-ends, *and* all the points along a single street segment where the street curves (see also Neis et al. 2011 for a discussion of other topological errors). The latter are not nodes in the graph-theoretic sense, so we remove them algorithmically and consolidate each resulting set of sub-edges between "true" network nodes (i.e., intersections and dead-ends) into single unified edges. These unified edges between intersections retain the full spatial geometry of the consolidated sub-edges and their relevant attributes, such as the length of the street segment.

OSMnx provides different simplification modes that offer fine-grained control to define nodes rigorously and reproducibly. In *strict* simplification mode, a node is either:

1. where an edge dead-ends, or
2. the point from which an edge self-loops, or
3. the intersection of multiple streets where at least one of the streets continues *through* the intersection (i.e., if two streets dead-end at the same point, creating an elbow, the point is not considered a node, but rather just a turn in a path)

In *non-strict* mode, the first two conditions remain the same, but the third is relaxed to allow nodes at the intersection of two streets, even if both streets dead-end there, as long as the streets have different OpenStreetMap IDs. A node is always retained at any point where a street changes from one-way to two-way.

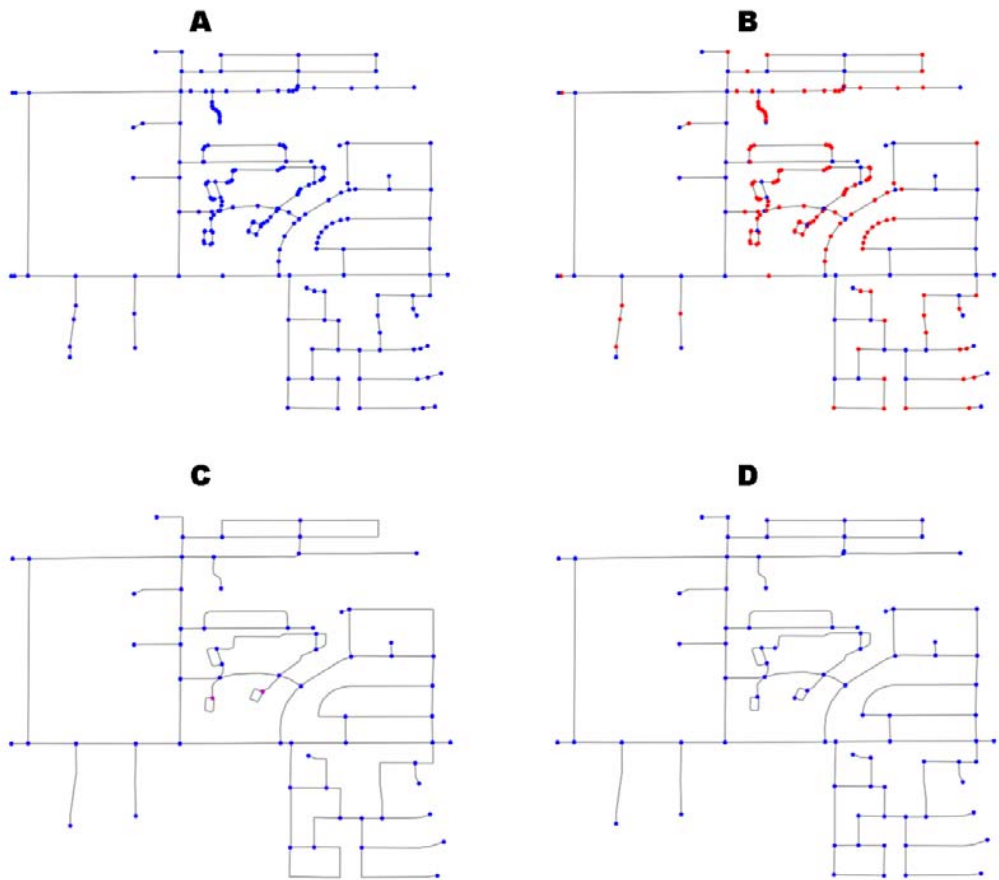

Figure 5. A) the original network, B) non-graph-theoretic nodes highlighted in red and true nodes in blue, C) strictly simplified network, with self-loops noted in magenta, D) non-strictly simplified network.

Figure 5 illustrates the topological simplification process. When we first download and assemble the street network from OpenStreetMap, it appears as depicted in Figure 5a. Then OSMnx automatically simplifies this network to only retain the nodes that represent the junction of multiple routes. First, it identifies all non-intersection nodes (i.e., all those that simply form an expansion graph), as depicted in Figure 5b. Then it removes them, but faithfully maintains the spatial geometry and attributes of the street segment between the true intersection nodes. In Figure 5c, all the non-intersection nodes have been removed, all the true nodes (dead-ends and junctions of multiple streets) remain in blue, and self-loop nodes are in magenta. In strict mode, OSMnx considers two-way intersections to be topologically identical to a single street that bends around a curve. Conversely, to retain these intersections when the incident edges have different OSM IDs, we use non-strict mode as depicted in Figure 5d.

### 3.4. Save street networks to disk

OSMnx can save a street network to disk as a GraphML file (an open, standard format for serializing graphs) to work with later in software such as Gephi or NetworkX. It can also save the network as shapefiles of nodes and edges to work with in any GIS. When saving as shapefiles, the network is simplified to an undirected representation, but one-way directionality and origin/destination nodes are preserved as edge attributes for GIS routing applications. OSMnx can also save street networks as scalable vector graphics (SVG) files for design work in Adobe Illustrator (Figure 6). OSMnx can also load preexisting GraphML files, so it is not limited to acquiring network data exclusively from OpenStreetMap.

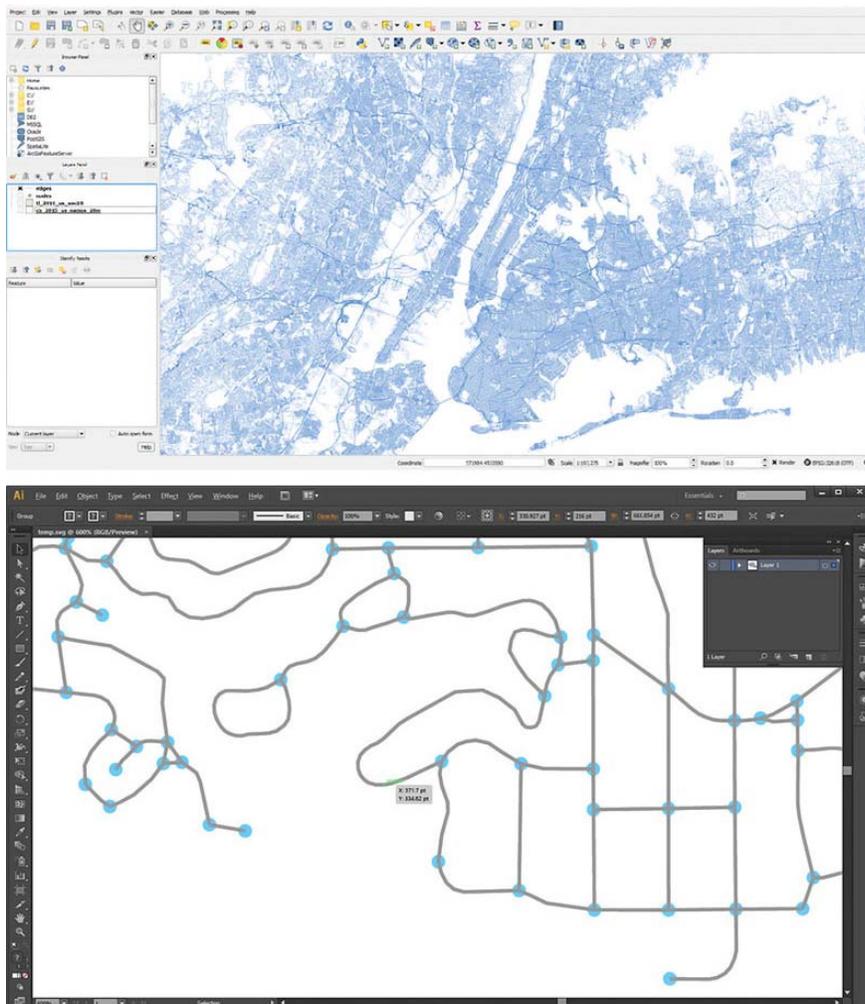

Figure 6. Street network for metropolitan New York from OSMnx saved and loaded in QGIS as a shapefile (above) and in Adobe Illustrator as SVG (below).

### 3.5. Analyze street networks

OSMnx analyzes networks and calculates network statistics, including spatial metrics based on geographic area or weighted by distance (Table 1). With a single command, OSMnx calculates the *individual nodes'* average neighborhood degrees (weighted and unweighted), betweenness centralities, closeness centralities, degree centralities, clustering coefficients (weighted and unweighted), PageRanks, and the *entire network's* intersection count, intersection density, average betweenness centrality, average closeness centrality, average degree centrality, eccentricity, diameter, radius, center, periphery, node connectivity, average node connectivity, edge connectivity, average circuity, edge density, total edge length, average edge length, average degree, number of edges, number of nodes, node density, maximum and minimum PageRank values and corresponding nodes, the proportion of edges that self-loop, linear street density, total street length, average street length, number of street segments, average number of street segments per node, and the counts and proportions of node types (i.e., dead-ends, 3-way intersections, 4-way intersections, etc.).

OSMnx can calculate and plot shortest-path routes between points or addresses, taking one-way streets into account (Figure 7). These shortest paths can be weighted by distance, travel time (assuming the availability of speed data), or any other impedance. For example, since OSMnx can automatically calculate street grades, shortest paths can minimize elevation change rather than trip distance. Moreover, the ability to calculate origin-destination distance matrices is built into NetworkX. OSMnx can also visualize street segments by length to provide a sense of where a city's longest and shortest blocks are distributed, or one-way versus two-way edges to provide a sense of where a city's one-way streets and divided roads are distributed. Researchers can quickly visualize the spatial distribution of dead-ends (or intersections of any type) in a city to get a sense of these points of network disconnectivity (Figure 8). Jacobs (1995) compared several cities' urban forms through figure-ground diagrams of one square mile of each's street network. We can re-create this automatically and computationally with OSMnx (Figure 9). These figure-ground diagrams are created completely with OSMnx and its network plotting function.

To summarize, OSMnx allows researchers and planners to download spatial geometries and construct, project, visualize, and analyze complex street networks. It automates the collection and computational analysis of street networks for powerful and consistent research, transportation engineering, and urban design. The following section illustrates its functionality with a simple case study.

| Measure | Definition |
|---|---|
| $n$ | number of nodes in network |
| $m$ | number of edges in network |
| average node degree | mean number of inbound and outbound edges incident to the nodes |
| intersection count | number of intersections in network |
| average streets per node | mean number of physical streets that emanate from each node (intersections and dead-ends) |
| counts of streets per node | a dictionary with keys = the number of streets emanating from the node, and values = the number of nodes with this number |
| proportions of streets per node | a dictionary, same as above, but represents a proportion of the total, rather than raw counts |
| total edge length | sum of edge lengths in network (meters) |
| average edge length | mean edge length in network (meters) |
| total street length | sum of edge lengths in undirected representation of network |
| average street length | mean edge length in undirected representation of network (meters) |
| count of street segments | number of edges in undirected representation of network |
| node density | $n$ divided by area in square kilometers |
| edge density | total edge length divided by area in square kilometers |
| street density | total street length divided by area in square kilometers |
| average circuity | total edge length divided by sum of great circle distances between the nodes incident to each edge |
| self-loop proportion | proportion of edges that have a single incident node (i.e., the edge links nodes $u$ and $v$, and $u=v$) |
| average neighborhood degree | mean degree of nodes in the neighborhood of each node |
| mean average neighborhood degree | mean of all average neighborhood degrees in network |
| average weighted neighborhood degree | mean degree of nodes in the neighborhood of each node, weighted by edge length |
| mean average weighted neighborhood degree | mean of all weighted average neighborhood degrees in network |
| degree centrality | fraction of nodes that each node is connected to |
| average degree centrality | mean of all degree centralities in network |
| clustering coefficient | extent to which node's neighborhood forms a complete graph |
| weighted clustering coefficient | extent to which node's neighborhood forms a complete graph, weighted by edge length |
| average weighted clustering coefficient | mean of weighted clustering coefficients of all nodes in network |
| PageRank | ranking of nodes based on structure of incoming edges |
| maximum PageRank | highest PageRank value of any node in the graph |
| maximum PageRank node | node with the maximum PageRank |
| minimum PageRank | lowest PageRank value of any node in the graph |
| minimum PageRank node | node with the minimum PageRank |
| node connectivity | minimum number of nodes that must be removed to disconnect network |
| average node connectivity | expected number of nodes that must be removed to disconnect randomly selected pair of non-adjacent nodes |
| edge connectivity | minimum number of edges that must be removed to disconnect network |
| eccentricity | for each node, the maximum distance from it to all other nodes, weighted by length |
| diameter | maximum eccentricity of any node in network |
| radius | minimum eccentricity of any node in network |
| center | set of all nodes whose eccentricity equals the radius |
| periphery | set of all nodes whose eccentricity equals the diameter |
| closeness centrality | for each node, the reciprocal of the sum of the distance from the node to all other nodes in the graph, weighted by length |
| average closeness centrality | mean of all the closeness centralities of all the nodes in network |
| betweenness centrality | for each node, the fraction of all shortest paths that pass through the node |
| average betweenness centrality | mean of all the betweenness centralities of all the nodes in network |

Table 1. Descriptions of network measures automatically calculated by OSMnx.

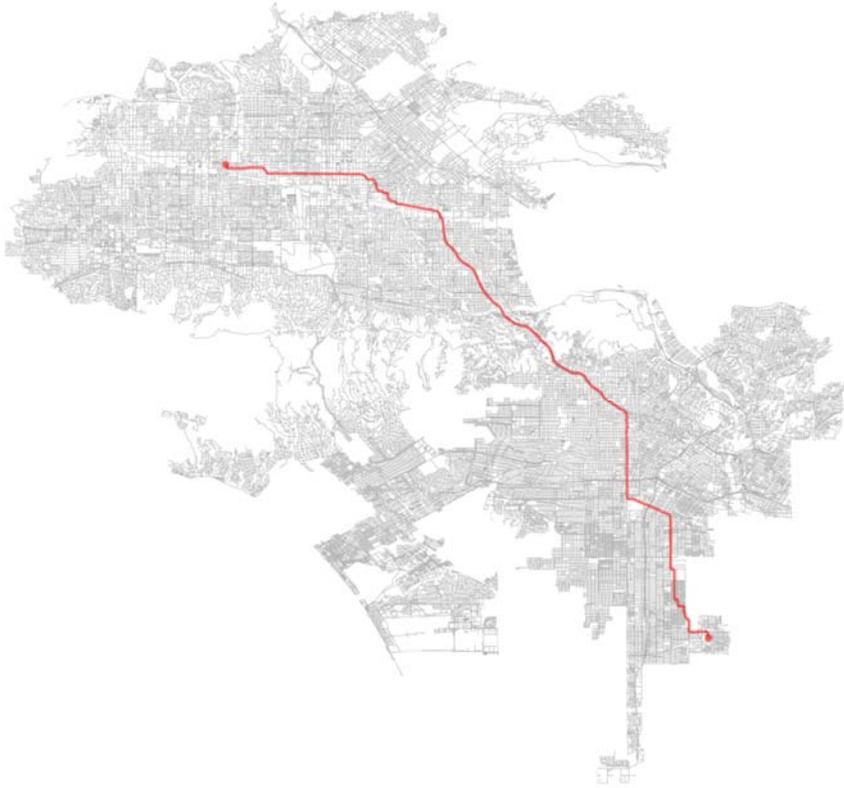

Figure 7. OSMnx calculates the shortest network path (weighted by edge length) between two points in Los Angeles, accounting for one-way routes, and plots it.

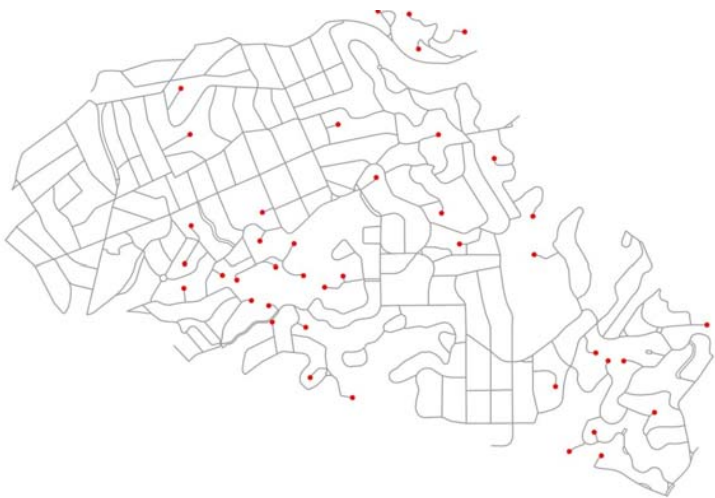

Figure 8. OSMnx visualizes the spatial distribution of dead-ends in Piedmont, California.

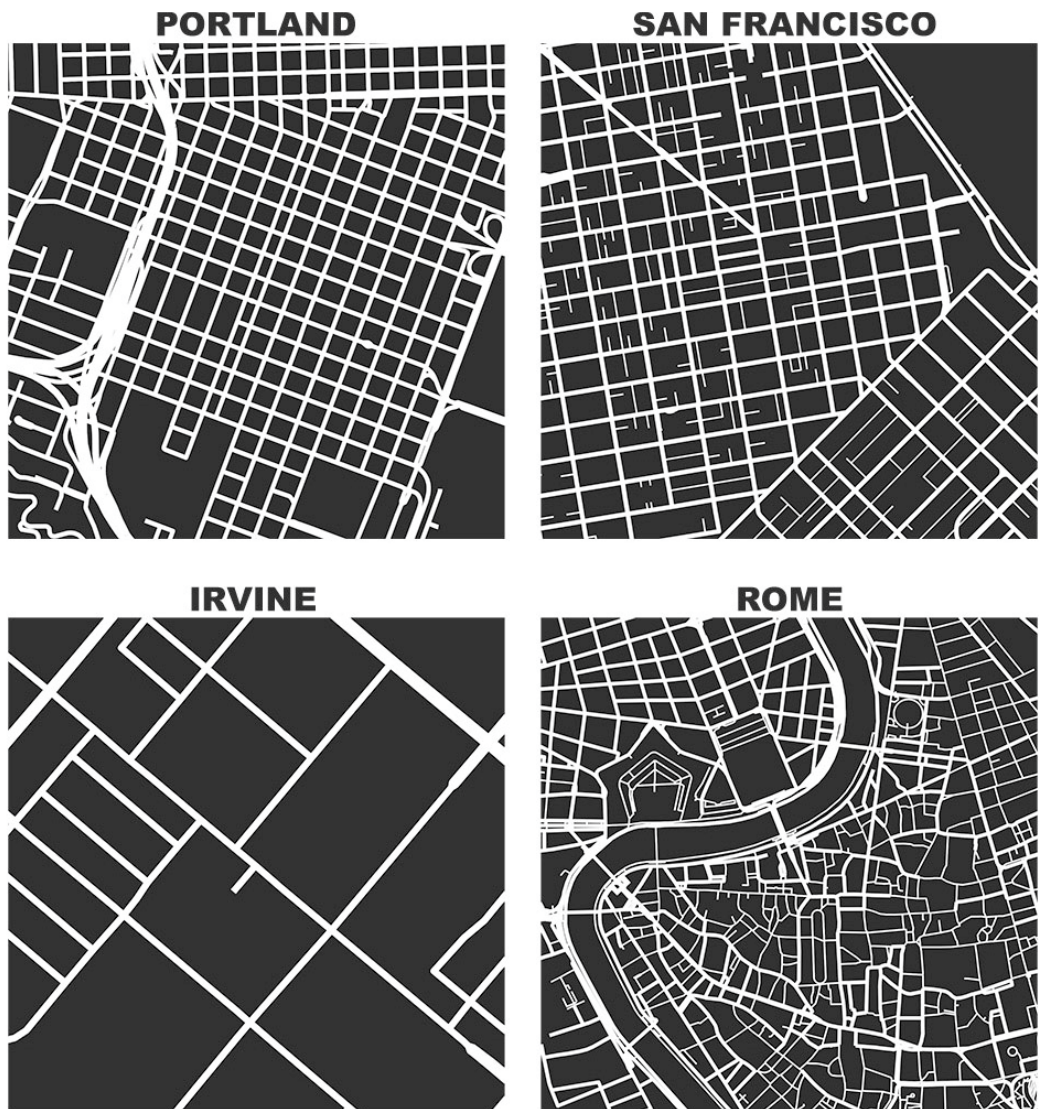

Figure 9. One square mile of each city, created and plotted automatically by OSMnx.

## 4. Case study: Portland, Oregon

To demonstrate OSMnx, we analyze three neighborhoods in Portland, Oregon. First, we define three square bounding boxes of 0.5 km$^2$ in the city's Downtown, Laurelhurst, and Northwest Heights neighborhoods. These study sites are small and do not conform to complete definitions of the local neighborhood boundaries, but are useful for visual comparison across sites at a small spatial scale (Snyder 1979). Next, OSMnx downloads the drivable, directed street networks for each, projects the networks to UTM (zone 10 calculated automatically), and plots them as seen in Figure 10. OSMnx calculates correct numbers of streets emanating from each node, even for peripheral intersections whose streets were cut off by the bounding box. In the network of Northwest Heights in Figure 10, some nodes appear to exist in the middle of a street segment, and thus should have been removed during simplification. However, these are actual intersections that OSMnx properly retained: they simply intersect a street that connects to a node outside the edge of the bounding box.

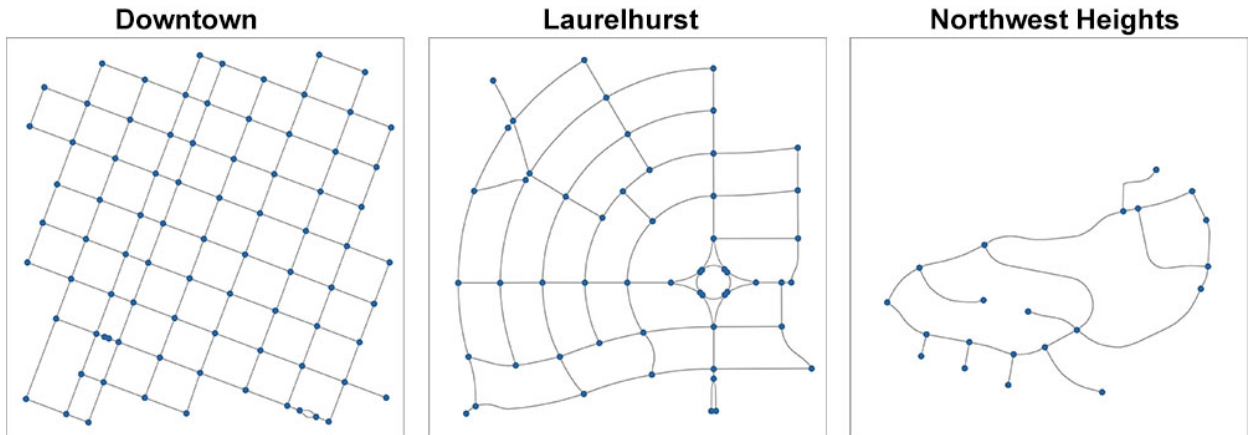

Figure 10. Three 0.5 km² sections of the street network in Portland, Oregon.

The different histories and designs of these street networks represent different historical eras, planning regimes, design paradigms, and topographies (Guzowski 1990; Grammenos and Pollard 2009; Works 2016). Several quantitative measures can describe these differences as well (Table 2). In terms of density metrics, Downtown has 164 intersections/km², Laurelhurst has 110, and Northwest Heights has 28. Downtown has 21 linear km of physical street/km², Laurelhurst has 16, and Northwest Heights has 5. In the Downtown network, the total street length *equals* the total edge length, because every edge is one-way. This differs in the other two networks because of the presence of two-way streets. As a proxy for block size, the average street segment length is 76 m in downtown, 92 m in Laurelhurst, and 117 m in Northwest Heights. These metrics quantify what we can see qualitatively by visually inspecting the street networks: Downtown's is fine-grained and dense, Laurelhurst's is moderate, and Northwest Heights' is coarse-grained and sparse.

|  | Downtown | Laurelhurst | NW Heights |
|---|---|---|---|
| Area (km²) | 0.5 | 0.5 | 0.5 |
| Avg of the avg neighborhood degree | 1.64 | 2.98 | 2.75 |
| Avg of the avg weighted neighborhood degree | 0.024 | 0.059 | 0.030 |
| Avg betweenness centrality | 0.070 | 0.077 | 0.137 |
| Avg circuity | 1.001 | 1.007 | 1.090 |
| Avg closeness centrality | 0.002 | 0.002 | 0.002 |
| Avg clustering coefficient | <0.001 | 0.108 | <0.001 |
| Avg weighted clustering coefficient | <0.001 | 0.023 | <0.001 |
| Intersection count | 82 | 55 | 14 |
| Avg degree centrality | 0.042 | 0.102 | 0.219 |
| Diameter (m) | 1278 | 1021 | 898 |
| Edge connectivity | 1 | 1 | 1 |
| Edge density (km/km²) | 21.32 | 29.55 | 10.71 |
| Avg edge length (m) | 76.3 | 97.4 | 116.6 |
| Total edge length (km) | 10.68 | 14.80 | 5.36 |
| Intersection density (per km²) | 163.7 | 109.8 | 28.0 |
| Average node degree | 3.42 | 5.53 | 4.38 |

| $m$ | 140 | 152 | 46 |
|---|---|---|---|
| $n$ | 82 | 55 | 21 |
| Node connectivity | 1 | 1 | 1 |
| Avg node connectivity | 1.326 | 2.107 | 1.443 |
| Avg node connectivity (undirected) | 2.868 | 2.496 | 1.443 |
| Node density (per km$^2$) | 163.7 | 109.8 | 41.9 |
| Max PageRank value | 0.030 | 0.029 | 0.106 |
| Min PageRank value | 0.002 | 0.004 | 0.017 |
| Radius (m) | 742.9 | 537.1 | 561.8 |
| Self-loop proportion | 0 | 0 | 0 |
| Street density (km/km$^2$) | 21.3 | 15.6 | 5.4 |
| Average street segment length (m) | 76.3 | 91.8 | 116.6 |
| Total street length (km) | 10.7 | 7.8 | 2.7 |
| Street segment count | 140 | 85 | 23 |
| Average streets per node | 3.93 | 3.58 | 2.38 |

Table 2. Descriptive statistics for three street network sections in Portland, Oregon.

Topological measures can tell us more about complexity, connectedness, and resilience. On average, nodes in Downtown have 3.9 streets emanating from them, in Laurelhurst they have 3.6, and in Northwest Heights they have 2.4. This is seen in Figure 10 as most intersections downtown are 4-way, whereas Laurelhurst features a mix of mostly 3-way and 4-way intersections, and Northwest Heights has mostly 3-way intersections and dead-ends. In fact, one-third of its nodes are the latter.

Unsurprisingly, as discussed earlier, the node and edge connectivity of each network is 1. More revealing is the average node connectivity. Recall that this represents the average number of nodes that must be removed to disconnect a randomly selected pair of non-adjacent nodes. In other words, this is how many non-overlapping paths exist on average between two randomly selected nodes. On average, 1.3 nodes must fail for two nodes to be disconnected in Downtown, 2.1 in Laurelhurst, and 1.4 in Northwest Heights (Table 2). These values may initially seem surprising: Downtown has the *least* resilient network despite its density and fine grain. However, this is explained by the fact that every edge in downtown is one-way, greatly circumscribing the number of paths between nodes. If we instead examine the *undirected* average node connectivity, it is 2.9 in Downtown, 2.5 in Laurelhurst, and 1.4 in Northwest Heights. Thus, were all edges in all three networks undirected, Downtown's average node connectivity would more than double and it would have the *most* robust network (by this measure). This finding suggests that there could be considerable complexity and resilience benefits in converting Downtown's streets from one-way to two-way – for the transportation modes that must obey directionality.

Finally, the average betweenness centrality indicates that 7% of all shortest paths pass through an average node in Downtown, 8% in Laurelhurst, and 14% in Northwest Heights. The spatial distribution of betweenness centralities in these three networks shows the relative importance of each node (Figure 11). In Downtown, important nodes are concentrated at the center of the network due to its grid-like orthogonality. In Northwest Heights, the two most important nodes are critical chokepoints connecting one side of the network to the other. In fact, the most important node in Northwest Heights has 43% of shortest paths running through it. By contrast, the most important node in Downtown has only 15%. The street network in this section of Northwest Heights is far more prone to disruption if its most important node fails (e.g., due to a traffic jam, flood, or earthquake) than Downtown is if its most important node fails.

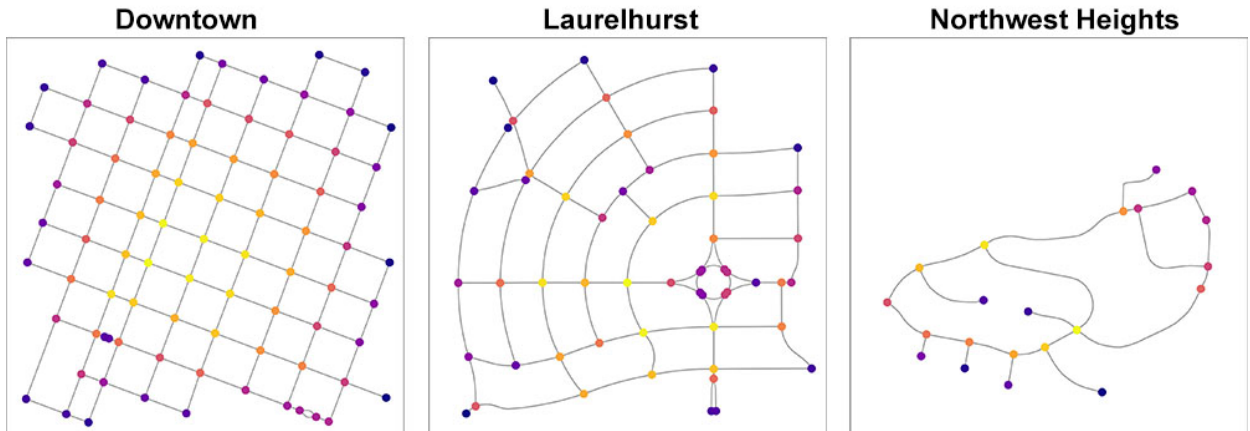

Figure 11. Three 0.5 km² sections of the street network in Portland, Oregon. Nodes colored by betweenness centrality from lowest (dark) to highest (light).

# 5. Discussion

Street network analysis currently suffers from challenges of usability, planarity, reproducibility, and sample sizes. This article presented a new open-source tool, OSMnx, to make the collection of data and creation and analysis of street networks easy, consistent, scalable, and automatable for any study site in the world. OSMnx contributes five capabilities for researchers and practitioners: downloading place boundaries and building footprints; downloading and constructing street networks from OpenStreetMap; correcting network topology; saving street networks to disk as ESRI shapefiles, GraphML, or SVG files; and analyzing street networks, including calculating routes, visualizing networks, and calculating metric and topological measures of the network. It addresses the usability and sample size challenges by enabling the easy acquisition and analysis of hundreds or thousands of street networks for anywhere in the world. This analysis natively blends graph-theoretic analysis with spatial analysis, and addresses the planarity challenge by using nonplanar directed graphs. Finally OSMnx enhances reproducibility by clearly defining spatial extents and topologies while serving as a free open source tool for anyone to re-run analyses.

In the simple case study of Portland, Oregon, we saw how to assess the street network from both metric and topological perspectives using OSMnx. The quantitative analysis corresponded with the qualitative assessment of the networks' visualizations. In particular, we found these networks differed substantially in density, connectedness, centrality, and resilience. As it is limited by its sample size, this small case study primarily serves illustrative purposes. Nevertheless, it demonstrates how to nearly instantaneously acquire, analyze, and visualize networks in just two or three lines of code with OSMnx. The small spatial scale of the analysis provides a succinct opportunity for clear visualization of the phenomena as well as neighborhood-scale interpretation of street network measures and their implications. However, these network subsets demonstrate peripheral edge effects in that they only consider flows *within* the subset, ignoring the rest of the city. A recent project addresses these limitations by using OSMnx to analyze 27,000 street networks at various scales across the United States (Boeing 2017; see appendix).

While these data provide various features about the built environment, they cannot tell us about the quality of the streetscape or pedestrian environment. OpenStreetMap is increasingly addressing this with richer attribute data about street width, lanes, speed limits, sidewalk presence, and street trees, but a general limitation of OSMnx is that it is dependent on what data exists in OpenStreetMap. While coverage is very good across the United States and Europe, developing countries have less thorough, but still quite adequate, street network coverage – especially in cities. Moreover, any researcher can digitize and add streets, building footprints, or other spatial data to OpenStreetMap at any time to serve as a public data repository for their own study, as well as anyone else's. In turn, OSMnx makes the acquisition, construction, and analysis of urban street networks easy, consistent, and reproducible while opening up a new world of public data to researchers and practitioners.

OSMnx is freely available on GitHub at https://github.com/gboeing/osmnx

# Appendix: Code and Data

This appendix demonstrates some simple code examples for working with OSMnx and describes a large repository of street network data and measures created with it. OSMnx itself is freely available online at https://github.com/gboeing/osmnx

### 6.1. Code examples

First we import OSMnx into the Python environment:

```
import osmnx as ox
```

Then we download, construct, correct, project, and plot the Los Angeles street network:

```
G = ox.graph_from_place('Los Angeles, California', network_type='drive')
ox.plot_graph(ox.project_graph(G))
```

If we want to get the walkable street network of Modena, Italy instead:

```
G = ox.graph_from_place('Modena, Italy', network_type='walk')
```

In the preceding examples, we use the graph_from_place function because OpenStreetMap has polygon boundaries for these places. However, it lacks polygon boundaries for some cities. In such cases, we can use the graph_from_address function to geocode the place name to a point, then get the network within some distance (in meters) of that point. If we want to get the street network of central Tunis:

```
G = ox.graph_from_address('Medina, Tunis, Tunisia', distance=1000)
```

We can similarly download networks by passing in shapefiles or latitude-longitude points. Once we have downloaded a network, we can calculate its measures:

```
stats = ox.basic_stats(G)
```

To download building footprints (wherever data is available):

```
buildings = ox.buildings_from_place('Piedmont, California')
```

### 6.2. Data repository

A recent study (Boeing 2017) used OSMnx to download and analyze 27,000 U.S. street networks at metropolitan, municipal, and neighborhood scales. This analysis comprises every U.S. city and town, census urbanized area, and Zillow-defined neighborhood. These street networks (shapefiles and GraphML files) and their measures have been shared in a public repository for other researchers to use at https://dataverse.harvard.edu/dataverse/osmnx-street-networks